\documentclass[aps,preprint,groupedaddress,showpacs]{revtex4}
\usepackage{amsmath}
\usepackage{epsfig}
\begin{document}

\preprint{HEP/123-qed}

\title[Short Title]{Bloch Equations and Completely Positive Maps}

\author{Sonja Daffer,$^1$
        Krzysztof W$\acute{\mbox{o}}$dkiewicz,$^{1,2}$}
\author{John K. McIver$^1$}%

\affiliation{%
    $^1$Department of Physics and Astronomy,
    University of New Mexico,
    800 Yale Blvd. NE,
    Albuquerque, NM 87131   USA  \\
    $^2$Instytut Fizyki Teoretycznej,
    Uniwersytet Warszawski, Ho$\dot{z}$a 69,
    Warszawa 00-681, Poland
    }%

\date{\today}      
\begin{abstract}
\vspace{.1in} \noindent The phenomenological dissipation of the
Bloch equations is reexamined in the context of completely
positive maps. Such maps occur if the dissipation arises from a
reduction of a unitary evolution of a system coupled to a
reservoir. In such a case the reduced dynamics for the system
alone will always yield completely positive maps of the density
operator. We show that, for Markovian Bloch maps, the requirement
of complete positivity imposes some Bloch inequalities on the
phenomenological damping constants. For non-Markovian Bloch maps
some kind of Bloch inequalities involving eigenvalues of the
damping basis can be established as well. As an illustration of
these general properties we use the depolarizing channel with
white and colored stochastic noise.
\\
\end{abstract}


\maketitle

\section{Introduction}
In 1946, Felix Bloch introduced a set of equations describing the
dynamics of a nuclear induction of a spin that interacts with a
magnetic field \cite{bloch1946}.  The applied magnetic field
drives the  Bloch vector of the magnetic moment, causing it to
precess about the field direction. In addition to the unitary
evolution describing the magnetic moment precession, a nonunitary
evolution is observed in nuclear magnetic resonance, which results
in dissipation of the magnetic observables. This dissipation is
characterized by two phenomenological decay constants. The two
lifetimes $T_1$ and $T_2$ are the longitudinal and transverse
decay constants, respectively.

Fluctuations in the environment, such as inhomogeneities in the
magnetic field and interactions with other moments, lead to
dissipation in the system. The dissipation constants $T_1$ and
$T_2$ are nonnegative so that exponential decay of the magnetic
moment  occurs. It is well known that this condition is required
to preserve the positivity of the density operator under
dissipation. The phenomenological dissipation of the Bloch vector
defines a positive map (PM) of the density operator for the spin
system.

About the physical sources of the dissipation constants Bloch
wrote:
\begin{quote}
{\it The actual value of $T_1$ is very difficult to predict for a
given substance ... To give a reliable estimate of $T_2$ ...
requires a more detailed investigation of the mechanism involved
and will not be attempted here.} \end{quote}
  In the same paper
Bloch made the following statement about the relative values of
the dissipation constants:
\begin{quote}
 {\it ... serious errors may be committed by
assuming $T_1= T_2$ . There are, on the other hand, also cases
where this equality is justified ...} \end{quote}

It took almost 30 years to understand that the dissipation results
from a reduction of a unitary evolution  of a system coupled to a
quantum reservoir. In the process of such a reduction, the
transformation of the density operator of the system has to be a
completely positive map (CPM). However, if the hypothesis of
complete positivity is to be imposed then the values of the
dissipation constants cannot be arbitrary. In particular, the
inequality $2 T_1 \geq T_2$ must hold and has been experimentally
observed \cite{abragam}.

It is the purpose of this paper to reexamine the well-known Bloch
equations in the context of completely positive maps.  We show
that the condition of complete positivity for Markovian Bloch maps
imposes some Bloch inequalities on the phenomenological damping
constants. For non-Markovian Bloch maps, generalized  Bloch
inequalities involving eigenvalues of the damping basis can be
established as well.  The depolarizing channel with white noise is
used to illustrate  these general properties. The non-Markovian
Bloch map is studied in the framework of a depolarizing channel
with colored noise.

\section{Bloch equations}
Although originally introduced in the context of nuclear magnetic
resonance, the Bloch equations are well-known in quantum optics,
where they describe a two-level atom interacting with an
electromagnetic field \cite{AE75}.  The Bloch equations offer a
physical picture of the density operator.  The dynamics of any
two-level quantum system can be expressed in terms of a
three-dimensional vector $\vec{b}=(u,v,w)^T$, called the Bloch
vector. For such systems, the set of all density operators can be
geometrically represented by a sphere with unit radius. States
that are on the surface of this Bloch sphere are pure states or
rank one density operators $\rho=| \psi \rangle \langle \psi |$.
States that are within the Bloch ball are mixed states, which are
written as convex combinations of pure states.

The optical Bloch equations are a set of differential equations,
one for each component of the Bloch vector, having the form
\begin{eqnarray}  \label{eq:bloch}
 \dot{u}&=&-\frac{1}{T_u} u-\Delta v,               \nonumber\\
 \dot{v}&=&-\frac{1}{T_v} v+\Delta u + \Omega w,      \\
 \dot{w}&=&-\frac{1}{T_w} (w-w_{eq})- \Omega v.      \nonumber
\end{eqnarray}
The unitary part of the evolution is governed by $\Omega$, the
Rabi frequency of the applied field.  The field is detuned from
the natural resonance of the atom by an amount $\Delta$. Note that
these  equations differ from the original Bloch equations  by the
fact that there are two different transverse dampings. These two
constants $T_u$ and $T_v$ are the decay rates of the in phase and
out of phase quadratures of the atomic dipole moment, while $T_w$
is the decay rate of the atomic inversion into an equilibrium
state $w_{eq}$. The interaction Hamiltonian of the system and
reservoir that leads to Eqs. (\ref{eq:bloch}) is
\begin{equation}   \label{intHam}
    H = \hbar (\Omega \sigma + \Omega^\star \sigma^\dagger )
    + \hbar [ \sigma \Gamma(t) + \sigma^\dagger \Gamma^\dagger(t)
    ],
\end{equation}
where $\sigma$ and $\sigma^\dagger$ are the lowering and raising
operators for the atomic system.  The master equation for the
entire system (S) and reservoir (R) is given by the von Neumann
equation ($\hbar=1$)
\begin{equation}
    \dot{\rho}_{SR}=-i [ H,\rho_{SR} ].
\end{equation}
The master equation for the system alone is obtained by tracing
over the environment degrees of freedom. The decay constants arise
due to an interaction with the reservoir  given by the variables
$\Gamma(t)$. This could, for example, be a collection of harmonic
oscillators, in which case $\Gamma(t)=\Sigma_k g_k b^\dagger_k
e^{-i(\omega-\nu_k)t}.$ In contrast to a quantum reservoir,
$\Gamma(t)$ could describe a classically fluctuating environment.

Typically, the phenomenological decay rates in the Bloch equations
appear as
\begin{equation}  \label{eq:decayrate}
   \frac{1}{T_u} = \frac{1}{T_2}, \hspace{.3in}
   \frac{1}{T_v} = \frac{1}{T_2}, \hspace{.3in}
   \frac{1}{T_w} = \frac{1}{T_1}.
\end{equation}
The damping of the component that is in phase with the driving
field is equal to the damping of the component that is out of
phase with the driving field.  The damping is caused by the
interaction of the system with an external environment that is
averaged over. This environment could be the vacuum field, which
leads to spontaneous emission.  Other phenomena, which result in
Eq. (\ref{eq:decayrate}), are coupling to a thermal field or phase
randomization due to atomic collisions.  The source of noise and
dissipation in the system is due to the fluctuation of the
environment.

The phenomenological decay rates are not limited to Eq.
(\ref{eq:decayrate}).   Reservoirs that result in different
dynamics can be engineered. For example, a two-level atom
interacting with a squeezed vacuum reservoir will experience
unequal damping for the in phase and out of phase quadratures of
the atomic dipole.  The corresponding damping rates become
\begin{equation}      \label{eq:svdamping}
  \frac{1}{T_u}=\frac{1}{T_2}+\frac{1}{T_3}, \hspace{.3in}
  \frac{1}{T_v}=\frac{1}{T_2}-\frac{1}{T_3}, \hspace{.3in}
  \frac{1}{T_w}=\frac{1}{T_1}.
\end{equation}
The presence of the parameter $T_3$ is the source of the damping
asymmetry between the $u$ and $v$ components of the Bloch vector.
It arises because the vacuum is squeezed, meaning that it has
fluctuations in one quadrature smaller than allowed by the
uncertainty principle at the expense of larger fluctuations in the
other quadrature. Equation (\ref{eq:svdamping}) reflects this --
the decay rate for one component of the Bloch vector is increased,
while the decay rate for the orthogonal component is
correspondingly decreased. The physical parameters leading to this
dynamics are
\begin{eqnarray}
\label{eq:squeezing}
 \frac{1}{T_u} &=& A \left( N+\frac{1}{2} +|M|\right),   \hspace{.1in}
 \frac{1}{T_v} = A \left( N+\frac{1}{2}-|M| \right), \\
 \frac{1}{T_w} &=& 2A \left( N+\frac{1}{2} \right),              \hspace{.1in}
 w_{eq} = -\frac{1}{2N+1},  \nonumber
\end{eqnarray}
where $A$ is the Einstein coefficient for spontaneous emission,
$N$ is the mean photon number of the squeezed vacuum reservoir,
and $M$ is the amount of squeezing of the reservoir.  The
parameter $N$ is related to the two-time correlation function for
the noise operators of the reservoir $\langle a^\dagger(t) a(t')
\rangle = N \delta (t-t')$ where $a(t)$ is the field amplitude for
a reservoir mode.  The squeezing complex parameter $M$ arises from
the two-time correlation function involving the square of the
field amplitudes $\langle a (t) a(t') \rangle = M^\star \delta
(t-t')$. The damping asymmetry parameter $ \frac{1}{T_3} =A |M|$
is in this case  due entirely to  squeezing. These are the
relations obeyed by squeezed white noise, which lead to squeezing
of a vacuum reservoir~\cite{gardiner1991}.

The solution to the Bloch equations determines the state of the
system for all time. The Bloch vector evolves in time according to
a linear map.  This linear map may be written in the form
\begin{equation}
  \Phi : \vec{b} \mapsto \vec{b'} = {\bf \Lambda} \vec{b} +
  \vec{t},
\end{equation}
where ${\bf \Lambda}$ is a damping matrix and $\vec{t}$ is a
translation.  The overall operation consists of contractions and
translations.  Due to the presence of translations, the
transformation is affine.

The damping matrix is a $3 \times 3$ matrix that takes the
diagonal form
\begin{equation}
  {\bf \Lambda} =
  \left(
  \begin{array}{ccc}
     \Lambda_1       &    0           &     0         \\
         0           &  \Lambda_2     &     0         \\
         0           &    0           &   \Lambda_3
  \end{array}
  \right).          \label{eq:dampingmatrix}
\end{equation}
We will show in Section V, that these eigenvalues can be
calculated using the damping basis for an appropriate master
equation describing the unitary and the dissipative dynamics of
the spin system.

Due to the correspondence between the Bloch vector $\vec{b}$ and
the density operator $\rho$, the linear map is a superoperator
that maps density operators into density operators according to
\begin{equation}
    \Phi : \rho \mapsto \rho'\,.
\end{equation}
The density operator can be expanded in the Pauli basis
$\sigma_{\alpha}=\lbrace I,\sigma_x,\sigma_y,\sigma_z \rbrace$ and
the components of $\rho$ transform under the map. This
transformation is characterized by a $4 \times 4$ matrix
representation of $\Phi$. It has been found that the general form
of any stochastic map on the set of complex $2 \times 2$ matrices
may be represented by a $4 \times 4$ matrix containing 12
parameters~\cite{kw1,ruskai2002}.

Without loss of generality, this $4 \times 4$ matrix may be cast
into the form
\begin{equation}    \label{eq:tdmatrixsvc}
  {\cal T}=
  \left(
  \begin{array}{cc}
    1              &         0         \\
    \vec{t}        &     \bf{\Lambda}
  \end{array}
  \right),
\end{equation}
which uniquely determines the map.  Hermiticity of the density
operator is preserved by requiring that ${\cal T}$ be real. The
first row must be $\lbrace 1,0,0,0 \rbrace$ to preserve the trace
of the density operator.

The matrix representation of the Bloch vector as an expansion in
terms of the Pauli matrices
\begin{equation}
  B=\vec{b} \cdot \vec{\sigma} =
  \left(
  \begin{array}{cc}
    w         &        u-iv  \\
    u+iv      &        -w
  \end{array}
  \right)
\end{equation}
illustrates the properties required by the linear map. In the
absence of noise, the Bloch vector remains on the Bloch sphere so
that
\begin{equation}
   \textrm{det} B = - (u^2 + v^2 +w^2)
\end{equation}
has magnitude unity. A general map transforms the matrix $B$
according to
\begin{equation}
   \Phi:B \mapsto B'.
\end{equation}
To guarantee that the map $\Phi$ transforms the density operator
into another density operator, the Bloch vector can be transformed
only into a vector contained in the interior of the Bloch sphere,
or the Bloch ball.  This requirement implies
\begin{equation}
   |\textrm{det} B'| \leq |\textrm{det} B|,
\end{equation}
so that the qubit density operator
\begin{equation}    \label{eq:blochrep}
   \rho=\frac{1}{2} \left( I+\vec{b} \cdot \vec{\sigma} \right) =
   \frac{1}{2}
   \left( I + B \right)
\end{equation}
under the map becomes
\begin{equation}
   \Phi(\rho)  : \rho \rightarrow \Phi (\rho) =\frac{1}{2} \left( I + B'
   \right).
\end{equation}
This is only possible if the $\Lambda_i$ in the damping matrix
contain contractions.  This is achieved in general for
$|\Lambda_i| \leq 1$. If $\Lambda_i=e^{-t/T_i}$ then $1/T_i \geq
0$ for $i=u,v,w$ is necessary for $\Phi$ to be a positive map;
\textit{i.e.}, it always maps positive operators into positive
operators.

The set of all pure states lie on the surface of the Bloch sphere
$u^2+v^2+w^2=1$. The map $\Phi$ takes this set into a set of
states that lie on the surface of an ellipsoid
\begin{equation}
  \left( \frac{u-t_{1}}{\Lambda_1} \right) ^2 +
  \left( \frac{v-t_{2}}{\Lambda_2} \right) ^2 +
  \left(  \frac{w-t_{3}}{\Lambda_3} \right)^2=1.
\end{equation}
Thus, it typically maps pure states into mixed states.  All
ellipsoids that are on or inside the Bloch sphere represent sets
of positive operators.  However, not all ellipsoids on or inside
the Bloch sphere correspond to completely positive dynamics.
\section{Completely positive maps}
The map $\Phi$ should be a completely positive map
\cite{kraus1983}. A completely positive map is defined by
\begin{equation}
\label{eq:CPM}
    \Phi \otimes I_n \geq 0
    \hspace{.05in} \forall \hspace{.05in} n \in \textrm{Z}_+,
\end{equation}
where the index $n$ is a positive integer in the set of all
positive integers $\textrm{Z}_+$.  The definition states that if
the dynamics $\Phi$ occurs on the system and external systems are
attached to the system, which evolve according to the identity
superoperator $I$, and if the the overall state is positive then
the map $\Phi$ is completely positive.

A completely positive map is required to describe reduced dynamics
because it implies that the reduced dynamics arises from a unitary
evolution
\begin{equation}   \label{eq:unitary}
    \Phi_t(\rho)=\textrm{Tr}_\Gamma \{ U(t) ( \rho \otimes
    |\gamma_0\rangle \langle \gamma_0|)U^\dagger(t) \},
\end{equation}
on the larger Hilbert space consisting of the system and the
environment. The environment degrees of freedom are denoted by
$\Gamma$ and $\gamma_0$ is some initial state of the environment.
Starting with a Hamiltonian for the closed system plus
environment, and then tracing or averaging over the environment
degrees of freedom, will always yield completely positive reduced
dynamics for the system alone.  Because of this requirement, there
are points inside the Bloch sphere that are not accessible.

If the reduced dynamics is consistent with Eq. (\ref{eq:unitary})
then it also has as a Kraus decomposition.  This implies the
existence of a set of operators $K_i$, called Kraus operators,
such that the map can be expressed as
\begin{equation}
  \Phi(\rho) = \sum_i K_{i}^{\dagger} \rho K_i,
  \label{eq:kraussvc}
\end{equation}
where the condition
\begin{equation}
    \sum_i K_i K^\dagger_i=I
    \label{eq:identity}
\end{equation}
ensures that unit trace is preserved for all
time~\cite{kraus1983}. If an operation has a Kraus decomposition,
then it is completely positive. The converse is also true.

To check whether a map $\Phi$ that takes $n \times n$ matrices
into $n \times n$ matrices is completely positive, it is necessary
and sufficient to check the positivity on a maximally entangled
$n^2 \times n^2$ state \cite{choi1972}.  This is a powerful
theorem that provides a test on a finite space, rather than
relying on the less practical definition, which requires attaching
systems in a countably infinite space.  In addition, it makes no
reference to Kraus operators, but rather, guarantees their
existence.

\section{Bloch inequalites}
A general completely positive, trace-preserving map for two-level
systems can always be written using four or fewer Kraus operators.
An important class of maps called unital maps (no translations)
has the following set of Kraus operators:
\begin{equation}
\begin{split}
    K_0 &= \frac{1}{2} \sqrt{1+\Lambda_1+\Lambda_2+\Lambda_3}\; I \\
    K_1 &= \frac{1}{2} \sqrt{1+\Lambda_1-\Lambda_2-\Lambda_3}\; \sigma_1 \\
    K_2 &= \frac{1}{2} \sqrt{1-\Lambda_1+\Lambda_2-\Lambda_3}\; \sigma_2 \\
    K_3 &= \frac{1}{2} \sqrt{1-\Lambda_1-\Lambda_2+\Lambda_3}\; \sigma_3.
\end{split}
\end{equation}
In order that the class of unital maps be completely positive, it
must be that the expressions under the radical are nonnegative.
The damping eigenvalues must obey the four inequalities
\begin{eqnarray}    \label{eq:inequalities}
  \Lambda_1 + \Lambda_2 - \Lambda_3 \leq 1,  \nonumber\\
  \Lambda_1 - \Lambda_2 + \Lambda_3 \leq 1, \\
  -\Lambda_1 + \Lambda_2 + \Lambda_3 \leq 1,  \nonumber  \\
  -\Lambda_1 - \Lambda_2 - \Lambda_3 \leq 1,    \nonumber
\end{eqnarray}
to guarantee complete positivity of the map.  This is a necessary
condition. This condition is more restrictive if compared with the
condition: $|\Lambda_i| \leq 1$ required by  a positive map
leading to contraction of the Bloch vector. In Figure 1 we have
depicted the completely positive maps as points inside of a
tetrahedron, forming a subset of all positive maps contained in a
cube. We shall call the relations (\ref{eq:inequalities}) Bloch
inequalities.


\begin{figure}[h]
\centering{
\includegraphics[angle=00,scale=0.60]{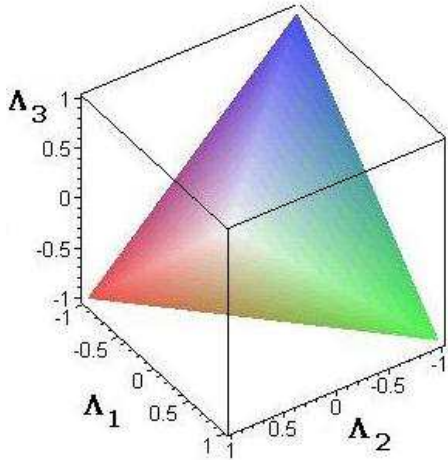}
}
\caption{\label{fig1a}%
Geometric representation of  PM versus CPM for the Bloch
equations. All points of the unit cube represent positive maps.
Points inside the tetrahedron are completely positive maps, given
by the conditions (\ref{eq:inequalities}).}
\end{figure}

The Bloch inequalities  (\ref{eq:inequalities}) lead to certain
restrictions on the damping constants in the Bloch equations.  For
purely exponential character of the $ \Lambda_i$, these conditions
lead to a simpler relation involving only lifetimes.  In this case
each component of the Bloch vector can only decay according to
\cite{gorini1976,kimura2002}:
\begin{equation}
\label{eq:blochineq}
\begin{split}
    \frac{1}{T_u} \leq \frac{1}{T_v} + \frac{1}{T_w}  \\
    \frac{1}{T_w} \leq \frac{1}{T_u} + \frac{1}{T_v}  \\
    \frac{1}{T_v} \leq \frac{1}{T_w} + \frac{1}{T_u}.
\end{split}
\end{equation}
The condition of complete positivity leads to a set of Bloch
inequalities for the phenomenological lifetimes that must be
satisfied. This explains the well-known phenomenon in nuclear
magnetic resonance whereby the inverse transverse relaxation time
is always less than or equal to twice the inverse longitudinal
relaxation time.

It is instructive to see how the violation of the Bloch
inequalities (\ref{eq:blochineq}) is reflected in the general
conditions (\ref{eq:inequalities}). In order to show this, we have
depicted in Fig. 2 the four conditions (\ref{eq:inequalities}) as
a function of time for two different selections of the damping
parameters. A function exceeding the straight line at the value
one is not a CPM, because it violates the Bloch inequalities. Note
that the case describe on the left figure is a PM that is not a
CPM for all times, while the case on the right figure describes a
CPM for all times.
\begin{figure}[h]
\centering{
\includegraphics[angle=00,scale=0.4]{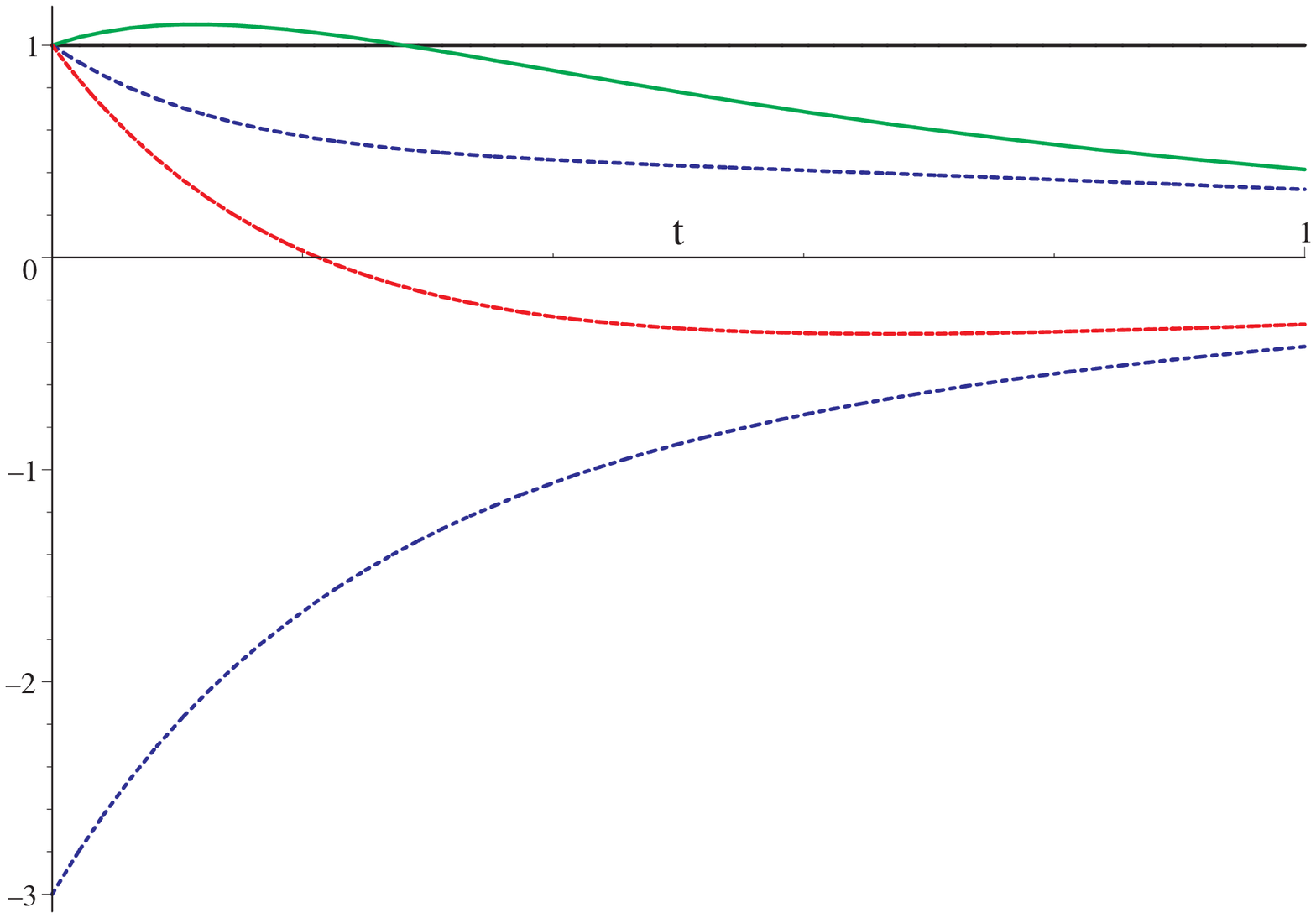}\hspace{1cm}
\includegraphics[angle=00,scale=0.4]{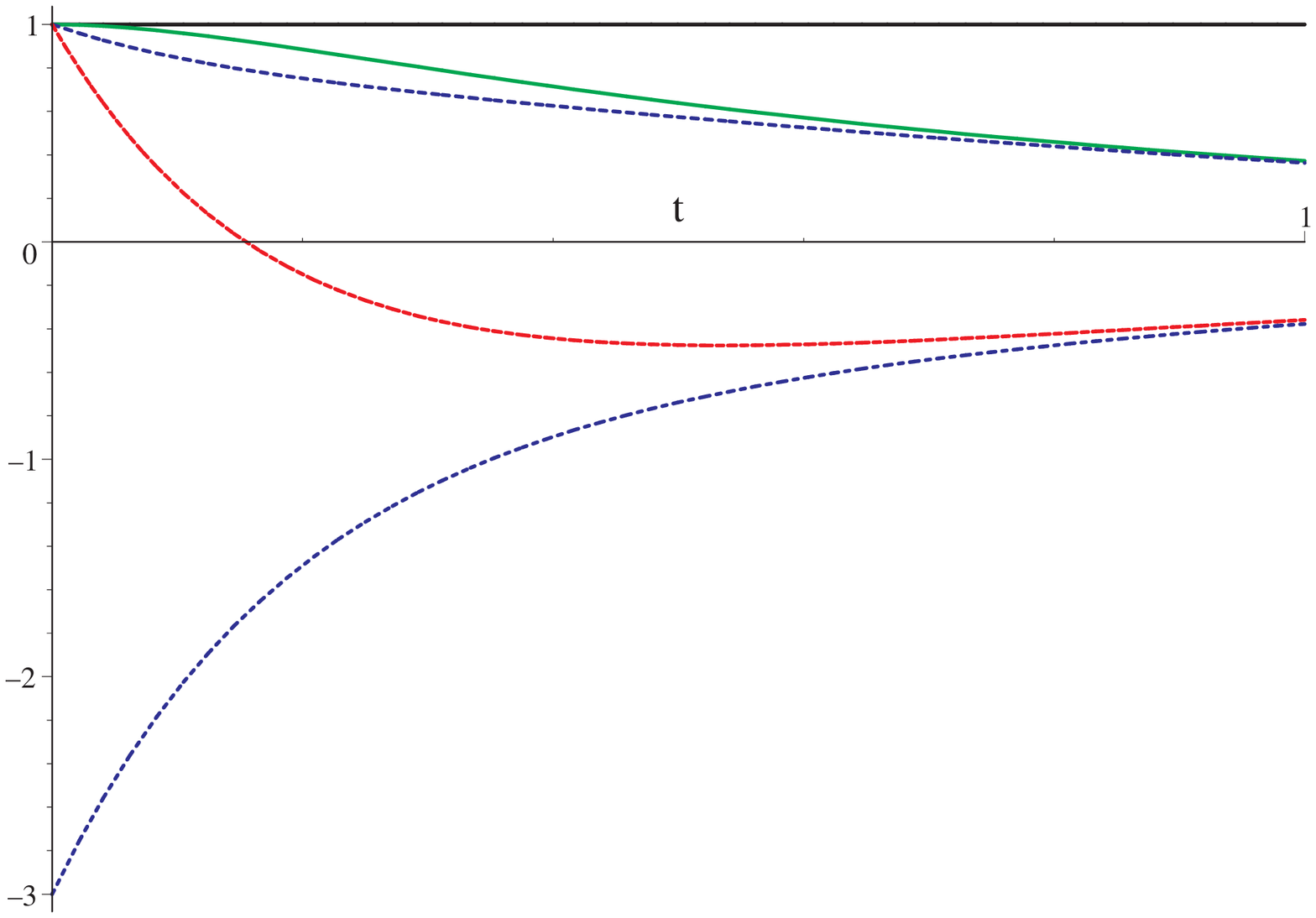}
} \vskip0.5cm
\caption{\label{fig2}%
Plots of the four conditions (\ref{eq:inequalities}) as a function
of time $t$. The left figure corresponds to $\frac{1}{T_u}=6,
\frac{1}{T_v}=3, \frac{1}{T_w}=1$. The right figure corresponds to
$\frac{1}{T_u}=6, \frac{1}{T_v}=5, \frac{1}{T_w}=1$. All values
are in arbitrary units. We see that the figure on the left is not
a CMP, because the Bloch inequalities (\ref{eq:blochineq}) are
violated.}
\end{figure}
Application of the Bloch inequalities  to the case of squeezed
noise given by Eqs. (\ref{eq:squeezing}) leads to the condition $
N+\frac{1}{2} \geq |M|$ \cite{gardiner1991}.

As a different example of a Bloch CMP generated by a noise, let us
consider the Hamiltonian
\begin{equation}  \label{eq:H3}
    H= x(t) \sigma_x + y(t) \sigma_y + z(t) \sigma_z,
\end{equation}
where $x(t)$, $y(t)$, and $z(t)$ are Gaussian random variables
with
\begin{equation}
\begin{split}
    \langle x(t) x(s) \rangle &=\frac{\Gamma_a}{2}
    \delta(t-s)\\
    \langle y(t) y(s) \rangle &=\frac{\Gamma_b}{2} \delta(t-s)\\
    \langle z(t) z(s) \rangle &=\frac{\Gamma_c}{2} \delta(t-s).
\end{split}
\end{equation}
Note that the white noise correlation has the form $\langle a(t)
a(s) \rangle = 2 D_a \delta(t-s)$, where the diffusion coefficient
is $D_a =\frac{\Gamma_a}{4}$. The same holds for the remaining
noises.

After averaging over the random variables, the Bloch vector will
transform by exponentiating the following matrix:
\begin{equation}    \label{eq:cpdampingmatrix}
  \left(
  \begin{array}{ccc}
         \Gamma_b+\Gamma_c       &     0         &     0         \\
         0             &    \Gamma_a+\Gamma_c    &     0         \\
         0             &     0         &     \Gamma_a+\Gamma_b
  \end{array}
  \right).
\end{equation}
As a result we obtain three different lifetimes given by the
following relations:

\begin{equation}
    \frac{1}{T_{u}} =\Gamma_b+ \Gamma_c,  \hspace{.1in}
    \frac{1}{T_{v}}=\Gamma_a+ \Gamma_c,  \hspace{.1in}
    \frac{1}{T_{w}} =\Gamma_a+ \Gamma_b.
\end{equation}
It is easy to verify that for positive diffusion coefficients
these lifetimes always satisfy the Bloch inequalities
(\ref{eq:blochineq}) i.e.,  are always generating completely
positive maps of the Bloch vector. An analysis of other cases in
which there are three phenomenological decay rates can be found in
Ref. \cite{daffer2003}.

\section{Markovian Bloch Map}
The Bloch equations with Gaussian white noise discussed in the
previous section are just an example of a general class of
Markovian completely positive maps. The general quantum Markovian
master equation for the system density operator can be written in
terms of the Gorini-Kossakowski-Sudarshan generator
\cite{gorini1976,kossakowski1972,lindblad1976}
\begin{equation}     \label{eq:gks}
     L \rho=-i [H,\rho ] + \mathcal{L} \rho,
\end{equation}
where the Lindblad superoperator may be written as
\begin{equation}     \label{eq:lin}
     \mathcal{L} \rho = \frac{1}{2} \sum^{n^2-1}_{i,j=1}
     c_{ij} \{ [F_i,\rho F^\dagger_j ] + [ F_i \rho, F^\dagger_j ]  \}
\end{equation}
for $\rho \in \mathcal{M}_n$ ($\mathcal{M}_n$ denotes the set of
$n \times n$ complex matrices), where $H=H^\dagger, \textrm{Tr}\{
H\}=0, \textrm{Tr}(F_i)=0, \textrm{Tr}\{ F^\dagger_i F_j
\}=\delta_{ij},$ and $(c_{ij})$ is a complex positive semidefinite
matrix.

The expression in Eq. (\ref{eq:gks}) has been shown to generate a
dynamical semigroup \cite{alicki1987} that has the following
properties:
\begin{eqnarray}   \label{eq:properties}
    &(i)& \parallel \Phi_t \rho \parallel_1 =
    \parallel \rho \parallel_1  \forall \hspace{.05in} \rho \in V_1^{+}(\mathcal{H}),
    \hspace{.05in} t\geq0,   \nonumber \\
    &(ii)& \Phi_t \otimes I_n \geq 0
    \hspace{.05in} \forall \hspace{.05in} n \in \textrm{Z}_+,  \\
    &(iii)& \textrm{lim}_{t \downarrow 0} \hspace{.05in} \Phi_t =
    I,     \nonumber\\
    &(iv)& \Phi_t \Phi_s = \Phi_{t+s}, \hspace{.05in} t,s\geq 0.
    \nonumber
\end{eqnarray}
Property $(i)$ states that the map is trace-preserving for all
positive operators, which form a positive cone
$V_1^{+}(\mathcal{H})$.  $\parallel \cdot \parallel_1$ denotes the
trace norm in the space of linear operators on the Hilbert space
$\mathcal{H}$.  The definition of complete positivity is given by
property $(ii)$. We have already encountered this property in Eq.
(\ref{eq:CPM}). The third property is a statement regarding
continuity. The map $\Phi$ is continuous from above and approaches
the identity superoperator. Property $(iv)$ is the essence of the
semigroup property.  It states that applying the map from time 0
to time $s$ and then applying the map from time $s$ to time $t$ is
equivalent to applying the map from time 0 to time $t+s$.
Obviously, exponential functions have this property.

In the following we will show that such Markovian maps can be
conveniently written in terms of a damping basis. The map $\Phi$
can be written in the general form
\begin{equation}
    \Phi (\bullet) = \sum_i \Lambda_i \textrm{Tr} \{ L^\dagger_i \bullet
    \}  R_i,
\end{equation}
where $L_i$ and $R_i$ are left and right eigenoperators, which
form a damping basis \cite{briegel993}. $L_i$ is the operator dual
to $R_i$ that satisfies the duality relation
\begin{equation}
 \textrm{Tr} \lbrace L_i R_j \rbrace = \delta_{ij}.
  \label{eq:duality}
\end{equation}
This is a complete, orthogonal basis with which to expand the
density operator at any time.

This basis is obtained by finding the eigenoperators of the
eigenvalue equation.  The right eigenoperators satisfy
\begin{equation}  \label{eq:righteigen}
  {\cal L} R_i = R_i \lambda,
\end{equation}
while the left eigenoperators satisfy the dual eigenvalue equation
\begin{equation}  \label{eq:lefteigen}
  L_i {\cal L} =\lambda L_i.
\end{equation}
Both $L_i$ and $R_i$ have the same eigenvalues.

If we use the eigenoperators $R_i$ of Eq. (\ref{eq:righteigen})
with corresponding eigenvalues $\lambda_i$, then once the initial
state is known
\begin{equation}
  \rho(0)= \sum_{i} \textrm{Tr} \lbrace L_i \rho(0) \rbrace R_i,
\end{equation}
the state of the system at any later time can be found through
\begin{eqnarray}    \label{eq:rhooft}
  \rho(t)=e^{{\cal L}t} \rho(0)
  =\sum_i \textrm{Tr} \lbrace L_i \rho(0) \rbrace \Lambda_i
  R_i   \\
  =\sum_i \textrm{Tr} \lbrace R_i \rho(0) \rbrace \Lambda_i
  L_i,   \nonumber
\end{eqnarray}
where $\Lambda_i$ are functions of both time and the eigenvalues
$\lambda_i$.  An arbitrary positive map must have the contractions
$|\Lambda_i| \leq 1$.

This method is a simple way of finding the density operator for a
given ${\cal L}$ for all times.  The solution of the left and
right eigenvalue equations yields a set of eigenvalues and
eigensolutions: $\lbrace \lambda_i, L_i , R_i \rbrace.$ Once the
damping basis is obtained, it can be used to expand the density
operator.

As an example let us investigate a stochastic model of a
depolarizing channel with white noise. As a simple example, we
will consider Bloch Eqs.(\ref{eq:bloch}) that have the following
phenomenological decay constants:
\begin{equation}   \label{eq:phasedecay}
 \frac{1}{T_1} = 0,                          \hspace{.2in}
 \frac{1}{T_2} = D,                          \hspace{.2in}
 \frac{1}{T_3} = 0,                          \hspace{.2in}
 w_{eq} = 0.
\end{equation}
This describes the process of phase randomization of the atomic
dipole caused by atomic collisions.  This leads to equal damping
for the $u$ and $v$ components of the Bloch vector with
contractions that depend on the parameter $D$ due to collisions.
Setting both $\Omega$ and $\Delta$ equal to zero, the Lindbladian
for this system is
\begin{equation}   \label{eq:depolarizingLindbladian}
    \mathcal{L} \bullet = - D [\sigma_3,[\sigma_3,
    \bullet]].
\end{equation}
$\mathcal{L}$ is the generator of the dissipative dynamics
\begin{equation}
    \Phi(\rho)= e^{t \mathcal{L}} \rho(0)
\end{equation}
and always generates a completely positive dynamical semigroup.
The decay constant is a scalar parameter that arises after tracing
or averaging over the environment degrees of freedom.  The
environment could be a quantum reservoir or a classically
fluctuating external system.  The von Neumann equation
\begin{equation}
    \dot{\rho}=-i z(t) [\sigma_3,\rho]
\end{equation}
for system and environment, containing a stochastic variable
$z(t)$, can be averaged over the noise exactly to obtain Eq.
(\ref{eq:depolarizingLindbladian}).  The noise leads to bit flip
errors in the system.

The damping basis for the bit flip Lindblad equation is given in
Pauli basis by
\begin{equation}
    R_\alpha = \frac{1}{\sqrt{2}} \sigma_\alpha\,.
\end{equation}
The left eigenoperators are identical to the right eigenoperators
because the Pauli operators are self-dual.  The damping
eigenvalues are given by
\begin{equation}
    \lambda_0=0, \hspace{.1in}\lambda_1=-4D, \hspace{.1in}
    \lambda_2=-4D, \hspace{.1in}\lambda_3=0.
\end{equation}
The time dependent functions $\Lambda_i$ become
\begin{equation}
    \Lambda_0=1, \hspace{.1in}\Lambda_1=e^{-4Dt}, \hspace{.1in}
    \Lambda_2=e^{-4Dt}, \hspace{.1in}\Lambda_3=1.
\end{equation}
The reduced dynamics describe dissipation of the system due to the
coupling of it with the environment.  The dissipation is in the
form of pure damping.  In the Bloch equations, this leads to
phenomenological decay constants
\begin{equation}
    \frac{1}{T_2}=\frac{1}{T_u}=\frac{1}{T_v}=4D, \hspace{.2in}
    \frac{1}{T_1}=\frac{1}{T_w}=0
\end{equation}
for each component of the Bloch vector.  The inversion, given by
the $w$ component undergoes no damping.  The two orthogonal
components $u$ and $v$ undergo equal damping.

The bit flip error equation results in a completely positive,
trace-preserving map.  Thus, there exists a set of Kraus
operators, which can be used to write the map as a decomposition
in terms of these operators.  Two Kraus operators can be used to
write the decomposition as
\begin{equation}
    \Phi(\rho)=K_0 \rho K^\dagger_0 + K_3 \rho K_3^\dagger,
\end{equation}
with Kraus operators explicitly given by
\begin{equation}
    K_0=\sqrt{\frac{1+e^{-4Dt}}{2}}\;I,
    K_3=\sqrt{\frac{1-e^{-4Dt}}{2}}\;\sigma_3.
\end{equation}
It is easy to show that these operators are normalized
\begin{equation}
    K_0^\dagger K_0 + K_3^\dagger K_3 =I,
\end{equation}
thus producing a trace-preserving completely positive map.

\section{Non-Markovian Bloch Map}
The damping basis method can be applied  to non-Markovian Bloch
maps. Such maps occur for example in a depolarizing channel with
colored noise. A bit flip error equation that does not rely on the
white noise assumption can be derived. This is achieved by
assuming the environment has correlations that are not of the form
of a delta-function, resulting in what is called colored noise.
The master equation is no longer of the Lindblad type; instead, it
contains a memory kernel operator leading to a master equation of
the general form
\begin{equation}
    \dot{\rho}=K \mathcal{L} \rho,
\end{equation}
where $K$ is an integral operator that depends on time of the form
$K \phi = \int_0^t k(t-t') \phi(t') \textrm{d}t'$. The kernel
function $k(t-t')$ is a well-behaved, continuous function that
determines the type of memory in the physical problem. The
solution to the master equation can be found by taking the Laplace
transform
\begin{equation}  \label{eq:laplace}
     s \tilde{\rho}(s) - \rho(0) = \tilde{K}(s) \mathcal{L}
    \tilde{\rho}(s),
\end{equation}
determining the poles, and inverting the equation in the standard
way.

To illustrate this class of master equations, the exponential
kernel function
\begin{equation}
    k(s,t)= e^{-\frac{|t-s|}{\tau}}
\end{equation}
is used. Consider the Hamiltonian $H=  z(t) \sigma_3$ with a
random telegraph signal (RTS) random variable $z(t)=a
(-1)^{n(t)}.$  The random variable $n(t)$ has a Poisson
distribution with a mean equal to $t/ 2\tau$, while $a$ is an
independent coin-flip random variable \cite{vankampen1981}.
 The equation of motion for the density operator is
given by the commutator
\begin{equation}
    \dot{\rho}=-i[H,\rho]=-i z(t)[\sigma_3,\rho].
\end{equation}

Taking the ensemble average over the random variables $z(t)$ leads
to an equation of motion for the average density operator
\cite{wodkiewicz1984}
\begin{equation}   \label{eq:eomforDC}
    \dot{\rho}=- a^2 \int_0^t e^{-\frac{t-s}{\tau}}[\sigma_3,[\sigma_3,\rho(s)]]\textrm{d}s.
\end{equation}
The brackets denoting the ensemble average are omitted and it
should be understood that we are considering the average density
operator. The master equation for the bit flip error with an
exponential memory kernel is exact.

The damping basis diagonalizes the equation.  This leads to a
single equation, for the two nontrivial components, that has the
following form:
\begin{equation}
    \frac{\textrm{d}}{\textrm{d}t} c(t) = -4 a^2 \int_0^t \textrm{d}s  e^{-\frac{|t-s|}{\tau}}
    c(s)  \textrm{d}s.
\end{equation}
Taking a derivative gives a second order differential equation
\begin{equation}
    \frac{\textrm{d}^2}{\textrm{d}t^2} c(t) +\frac{1}{\tau}  \frac{\textrm{d}}{\textrm{d}t} c(t)
    +4 a^2 c(t) = 0\,.
\end{equation}
The function $c(t)$ is the solution to a damped harmonic
oscillator equation of motion
\begin{equation}
    c(t)= e^{-\frac{t}{2 \tau}} \left(
    \cos (\Omega t)+\frac{\sin(\Omega t)}{2 \Omega \tau} \right),
\end{equation}
where $\Omega=\sqrt{4 a^2-\frac{1}{4 \tau^2}}$ and $c(0)=1$.

The image of this random telegraph signal (RTS) map is similar to
that of a depolarizing channel with white noise.  The
transformation given by Eq. (\ref{eq:tdmatrixsvc}) for this noise
is
\begin{equation}
  {\cal T}=
  \left(
  \begin{array}{cccc}
    1        &       0      &       0      &        0     \\
    0        &      c(t)    &       0      &        0     \\
    0        &       0      &      c(t)    &        0     \\
    0        &       0      &       0      &        1
  \end{array}
  \right),
\end{equation}
which has the same form as the depolarizing channel.  Notice that
there are no translations of the Bloch vector and
\begin{equation}
\label{lambda} \Lambda(t)=\Lambda_1= \Lambda_2=c(t),\ \
\Lambda_2=c(t), \ \ \mathrm{and} \ \  \Lambda_3=1.
\end{equation}
 The depolarizing channel with white noise has
simple contractions so that $c(t)$ is an exponential function.
This is a property of white noise.  The RTS channel has colored
noise leading to a nonexponential function for $\Lambda(t)$.
Rather, it contains oscillating terms with an exponential
envelope. A power series expansion gives

\begin{equation}
\Lambda(\nu)= 1-2a^2t^2+O(t)^3\,,
\end{equation}
 which
shows that the linear term in $t$ is missing. Thus, the standard
white noise diffusion term vanishes. This is in contrast to the
Markovian case, where the functions are purely exponential
functions in time with parameters defining the characteristic
lifetimes. This is a general property of the memory kernel and a
fundamental difference between white noise and colored noise. The
white noise limit can be recovered from (\ref{lambda}) with a
singular limit  of $\tau \rightarrow 0$ and $\gamma= 4a^2
=\mathrm{const}$. In such a white noise limit
$\Lambda_{\mathrm{wn}}=e^{-\gamma t}$.

 In terms of the dimensionless time $\nu=t/ 2\tau$
 we can write $\Lambda(\nu)=e^{-\nu}(\cos(\mu
\nu)+\frac{\sin{(\mu \nu)}}{\mu})$, with $\mu=\sqrt{(4 a
\tau)^2-1}$. The function $\Lambda(\nu)$ has two regimes -- pure
damping and damped oscillations. The fluctuation parameter, given
by the product $a \tau$, determines the behavior of the solution.
When $0 \leq a \tau < 1/4$ the solution is described by damping.
The frequency $\mu$ is imaginary with magnitude less than unity.
When $a \tau = 1/4$ the function $\Lambda(\nu)= e^{-\nu} (1-\nu)$
is unity at the initial time and approaches zero as time
approaches infinity. In addition to pure damping, damped harmonic
oscillations in the interval $[-1,+1]$ exist in the regime $a \tau
> 1/4$.

The Kraus operators for the RTS channel are similar to those for
the depolarizing channel.  The non-Markovian Bloch map can be
defined as
\begin{equation}
    \begin{split}
    \Phi(\rho)&=K_1^\dagger \rho K_1 +K_2^\dagger \rho K_2  \\
    &= \frac{1}{2}[ 1+\Lambda(t)]\, \rho +\frac{1}{2}[
    1-\Lambda(t) ]\, \sigma_3 \rho  \sigma_3.
    \end{split}
\end{equation}
The Bloch sphere evolves into an ellipsoid according to the
equation
\begin{equation}
    \left( \frac{u}{\Lambda(t)} \right)^2 + \left( \frac{v}{\Lambda(t)} \right)^2
    + w^2 = 1.
\end{equation}
  Unlike the exponential damped
solution for white noise, the function can take on negative values
and is bounded between $\pm 1$.

\section{Noise and separability}

The maximally entangled Bell state can be used to test if an
arbitrary map for qubits is completely positive.  The initial
entangled state for two qubits is given by a positive semidefinite
density operator.  If only one qubit of the joint state is
subjected to noise, the joint state must also be described by a
positive semidefinite density operator for all time.  For
$n$-level systems, positivity of the maximally entangled
$n^2$-level state implies that the map for the $n$-level part of
the state is completely positive.

The maximally entangled two qubit state can also be used to
address the issue of separability.  The partial transposition map
is an example of a map that is positive but not completely
positive. The partial transposition map applied to an entangled
$n^2$ state involves performing the transpose operation on one
half of the state, while performing the identity operation on the
other part of the state.  Note that the partial transposition map
is not a continuous map connected to the identity superoperator.
Although the initial entangled state is described by a positive
semidefinite density operator, the resulting transposed state can
be negative.  Using the Peres criterion of the positivity of the
partial transpose, one can determine when the state becomes
separable~\cite{peres1996}. A necessary and sufficient condition
for the output state to be nonseparable is that the partial
transpose map be negative~\cite{horodecki1996}. To check the
positivity of the partial transpose it suffices to examine the
eigenvalues of the operator given by
\begin{equation}   \label{eq:Perescondition}
    \widetilde{\Phi}(\rho_{AB})=\mathbf{1}_A \otimes \mathrm{T}_B
    [\Phi(\rho_{AB})]
\end{equation}
where $\mathrm{T}_B$ denotes the transpose of the state of Bob's
qubit. The dynamical map $\Phi$ is first applied to the maximally
entangled state in the following way: $\Phi(\rho_{AB})=I_A \otimes
\Phi_B [\rho_{AB}]$. The transposition map acts on one half of the
entangled state and the output state is given by
$\widetilde{\Phi}(\rho_{AB})$ in Eq. (\ref{eq:Perescondition}).

The four eigenvalues of the partial transpose matrix for the bit
flip error are
\begin{eqnarray}
\begin{split}
    e_1 = e_2 =\frac{1}{2}, \hspace{.2in}
    e_3 = - \frac{\Lambda(t)}{2}, \hspace{.2in}
    e_4 =\frac{\Lambda(t)}{2}\,.
\end{split}
\end{eqnarray}
>From this relation we conclude that such a state is separable if
and only if $\Lambda(t)=0$. This shows that in the limit of
Markovian dynamics, the separability is reached only in the
asymptotic limit of $t \rightarrow \infty$. The situation is
remarkably different for a non-Markovian map. In Figure 3 we have
depicted the separability condition as a function of the color of
the noise, characterized by the parameter $\tau$.

\begin{figure}[h]
\centering{
\includegraphics[angle=00,scale=0.40]{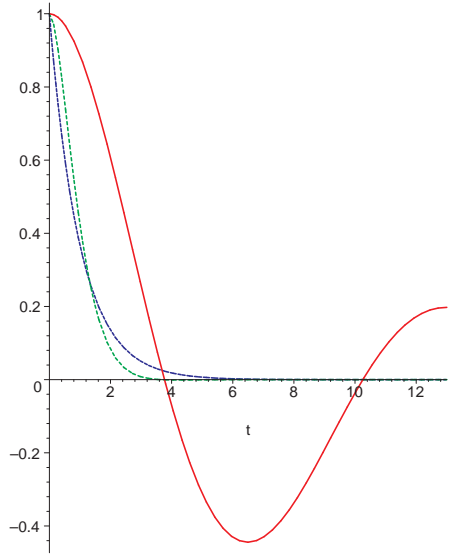}
}
\caption{\label{fig1b}%
Plots of non-Markovian $\Lambda(t)$  and $\tau = 4$, (solid line),
$\tau=0.3$ (dot-line), white noise  case $\tau=0$ (dash-line). All
plots in arbitrary units with  $\gamma=1$. The state becomes
separable for $\Lambda(t)=0$.}
\end{figure}

\section{Conclusion}
The Bloch equations are well-known to physicists working in the
field of nuclear resonance or quantum optics. These equations have
been widely used in quantum information theory to describe the
dissipation of various quantum channels for qubits. The original
Bloch equations have been derived on a purely phenomenological
ground, and phenomenological damping constants have been
introduced. We have shown that the physical consequences of a
quantum mechanical description of dissipation leads to complete
positivity of the Bloch maps. The condition that these maps are
CPM, implies a set of Bloch equations for the generalized
eigenvalues of the damping matrix. In the case of pure Markovian
dissipations these Bloch equations can be reduced to simple
inequalities for the various lifetimes characterizing the qubit.
We have shown that this approach can be extended to non-Markovian
dynamics. In this case the Bloch  inequalities involve
time-dependent eigenvalues of the generalized damping matrices. We
have illustrated our points with examples of Markovian and
non-Markovian depolarizing channels for qubits. We have shown a
fundamental difference in the separability properties of
correlated qubits in such channels.
\begin{acknowledgments}
This paper was been written to honor the 60 year birthday of Prof.
Rz\c{a}\.{z}ewski, whose work in the field of quantum optics is
well recognized. This work was partially supported by a KBN grant
No PBZ-Min-008/P03/03 and the European Commission through the
Research Training Network
QUEST.\\
\end{acknowledgments}











\begin{thebibliography}{}

\bibitem[Bloch (1946)] {bloch1946} F. Bloch, Phys. Rev {\bf 70},
460 (1946).


\bibitem[Abragan (19610]{abragam} A. Abragam, {\it Principles of
Nuclear Magnetism}, Oxford University Press, Oxford, (1961).

\bibitem[Allen Eberly (1975] {AE75} L. Allen and J. H. Eberly,
{\it Optical Resonance and Two-Level Atoms}, Wiley, New York,
(1975).



\bibitem{gardiner1991}  C.W. Gardiner, {\it
Quantum Noise}, Springer-Verlag, Berlin (1991).

\bibitem{kw1} K. W{\'odkiewicz}, Optics
Express {\bf 8}, No. 2,
 145 (2001).

\bibitem{ruskai2002}M.B. Ruskai, E. Werner, and S. Szarek,
 Linear Algebra Appl. {\bf 347}, 159 (May, 2002).




\bibitem[Kraus (1983)] {kraus1983}  K. Kraus, {\it States, Effects and
Operations: Fundamental Notions of Quantum Theory},
(Springer-Verlag, Berlin, 1983).

\bibitem[Choi (1972)] {choi1972}  M. Choi, Can. J.
Math. {\bf 24}, No. 3, 520 (Jan., 1972).


\bibitem[Gorini (1976)] {gorini1976}  V. Gorini, A. Kossakowski,
and E.C.G. Sudarshan, J. Math. Phys. {\bf 17}, No. 5, 821, (1976).

\bibitem[Kimura (2002)]{kimura2002} G. Kimura,
\pra {\bf 66}, 062113 (2002).


\bibitem[Daffer (2003)]{daffer2003} S. Daffer, K. Wodkiewicz, and
J.K. McIver, \pra {\bf 67}, 062312 (2003).


\bibitem[Kossakowski (1972)]{kossakowski1972} A. Kossakowski,
Bull. Acad. Polon. Sci., S\'{e}r. Sci. Math. Astronom. Phys. {\bf
20}, 1021 (1972).

\bibitem[Lindblad (1976)]{lindblad1976} G. Lindblad,
Commun. Math. Phys. {\bf 48}, 119 (1976); F. Lindblad, {\it
Non-Equilibrium Entropy and Irreversibility}, (Reidel, Dordrecht,
1983).

\bibitem[Briegel (1993)]{briegel993} H.J. Briegel and B.-G. Englert,
\pra {\bf 47}, 3311 (1993).

\bibitem[Alicki (1987)] {alicki1987}  R. Alicki and K. Lendi, {\it
Quantum Dynamical Semigroups and Applications}, (Springer-Verlag,
Berlin, 1987).

\bibitem[VanKampen (1981)] {vankampen1981} S.O. Rice, Bell Syst. Tech. J., {\bf 23}, 282
(1944); N.G. Van Kampen, {\it Stochastic Processes in Physics and
Chemistry}, (Elsevier, Amsterdam, 1992).

\bibitem[Wodkiewicz (1984)]{wodkiewicz1984} J.H. Eberly, K.
W\'{o}dkiewicz, and B.W. Shore, \pra {\bf 30}, No. 5, 2381 (1984).

\bibitem{peres1996}
A. Peres, Phys. Rev. A {\bf 77}, 1413 (1996).

\bibitem{horodecki1996}
M. Horodwecki, P. Horodecki and R. Horodecki, Phys. Lett. {\bf
A233},1 (1996).
\end{thebibliography}
\end{document}